\documentclass[twoside,a4paper,11pt]{proceedings}
\usepackage{graphicx}
\usepackage{hyperref}
\usepackage{color}
\usepackage{natbib}
\begin{document}
\pagenumbering{arabic}
\pagestyle{myheadings}
\thispagestyle{empty}
\vspace*{-1.5cm}
\parbox{\hsize}{European Week of Astronomy and Space Science (Athens, 4-8 July 2016)\\
                {\footnotesize EWASS Symposium~16: \emph{Frontiers of massive-star evolution and core-collapse supernovae}}}
\vspace*{0.2cm}
\begin{flushleft}
{\bf {\LARGE
The impact of bars on the radial distribution of supernovae in disc galaxies
}\\
\vspace*{1cm}
A.~A.~Hakobyan\footnote{{\footnotesize E-mail: hakobyan@bao.sci.am}},
A.~G.~Karapetyan$^{1}$,
L.~V.~Barkhudaryan$^{1}$,
G.~A.~Mamon$^{2}$,
D.~Kunth$^{2}$,
A.~R.~Petrosian$^{1}$,
V.~Adibekyan$^{3}$,
L.~S.~Aramyan$^{1}$
and M.~Turatto$^{4}$
%
}\\
\vspace*{0.5cm}
%
{\small
$^{1}$
Byurakan Astrophysical Observatory, Byurakan, Armenia\\
$^{2}$
Institut d'Astrophysique de Paris, Paris, France\\
$^{3}$
Instituto de Astrof\'{i}sica e Ci\^{e}ncia do Espa\c{c}o, Porto, Portugal\\
$^{4}$
Osservatorio Astronomico di Padova, Padova, Italy
}
%
\end{flushleft}
\markboth{
The impact of bars on the distribution of SNe
}{
A.~A.~Hakobyan~et~al.
}
\thispagestyle{empty}
\vspace*{0.4cm}
\begin{minipage}[l]{0.09\textwidth}
\
\end{minipage}
\begin{minipage}[r]{0.9\textwidth}
\vspace{0.1cm}
\section*{Abstract}{\small
We present an analysis of the impact of bars on the radial distributions
of the different types of supernovae (SNe) in the stellar discs of
host galaxies with various morphologies. We find that in Sa--Sbc galaxies,
the radial distribution of core-collapse (CC) SNe in barred hosts
is inconsistent with that in unbarred ones, while the distributions of
SNe Ia are not significantly different. At the same time, the radial
distributions of both types of SNe in Sc--Sm galaxies are not affected by bars.
We propose that the additional mechanism shaping the distributions of
Type Ia and CC SNe can be explained within the framework of substantial
suppression of massive star formation in the radial range swept by strong bars,
particularly in early-type spirals. The radial distribution of CC SNe in
unbarred Sa--Sbc galaxies is more centrally peaked and inconsistent with
that in unbarred Sc--Sm hosts, while the distribution of SNe Ia in unbarred
galaxies is not affected by host morphology. These results can be explained
by the distinct distributions of massive stars in the discs of
early- and late-type spirals.
\vspace{10mm}
\normalsize}
\end{minipage}

\section*{Summary}

This is a brief summary of \citet{2016MNRAS.456.2848H},
written for a short contribution in the EWASS-2016 Symposium~16
\emph{``Frontiers of massive-star evolution and core-collapse supernovae''}.
In the mentioned paper, using a well-defined and homogeneous
sample of SNe and their host galaxies from the coverage of
Sloan Digital Sky Survey \citep{2012A&A...544A..81H}, we analysed the impact
of bars on the radial distributions of the different types of SNe
in the stellar discs of host galaxies with various morphologies.
Our sample consists of 419 nearby (${\leq {\rm 100~Mpc}}$),
low-inclination ($i \leq 60^\circ$), and morphologically non-disturbed S0--Sm galaxies,
hosting 500 SNe in total.

All the results that we summarize below concerning the radial distributions of SNe
in barred galaxies can be explained considering the strong impact of the bars on
the distribution of massive star formation in stellar discs of galaxies,
particularly in early-type spirals.

\begin{figure}
\begin{center}$
\begin{array}{@{\hspace{0mm}}c@{\hspace{0mm}}c@{\hspace{0mm}}}
\includegraphics[width=0.5\hsize]{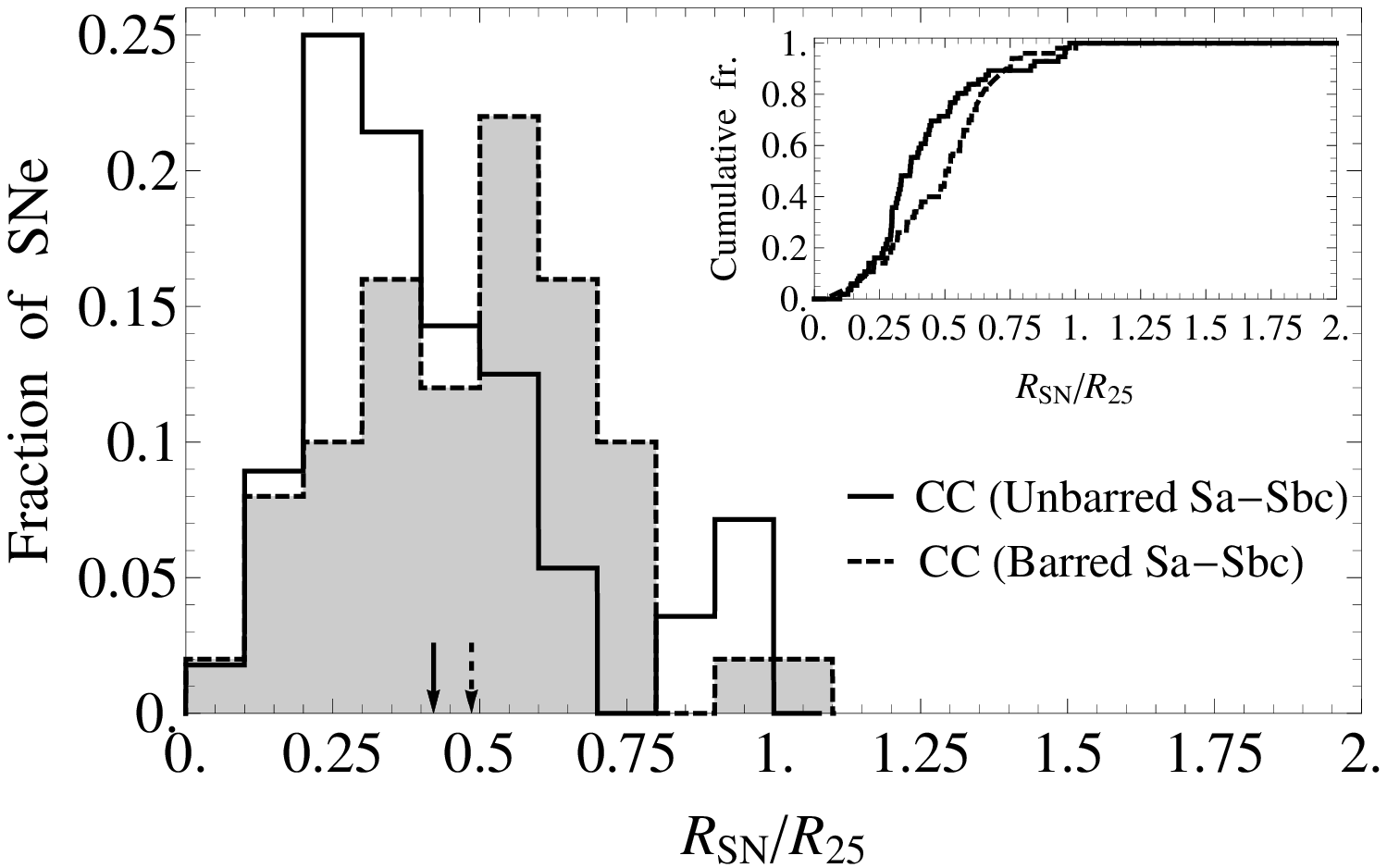}
\includegraphics[width=0.5\hsize]{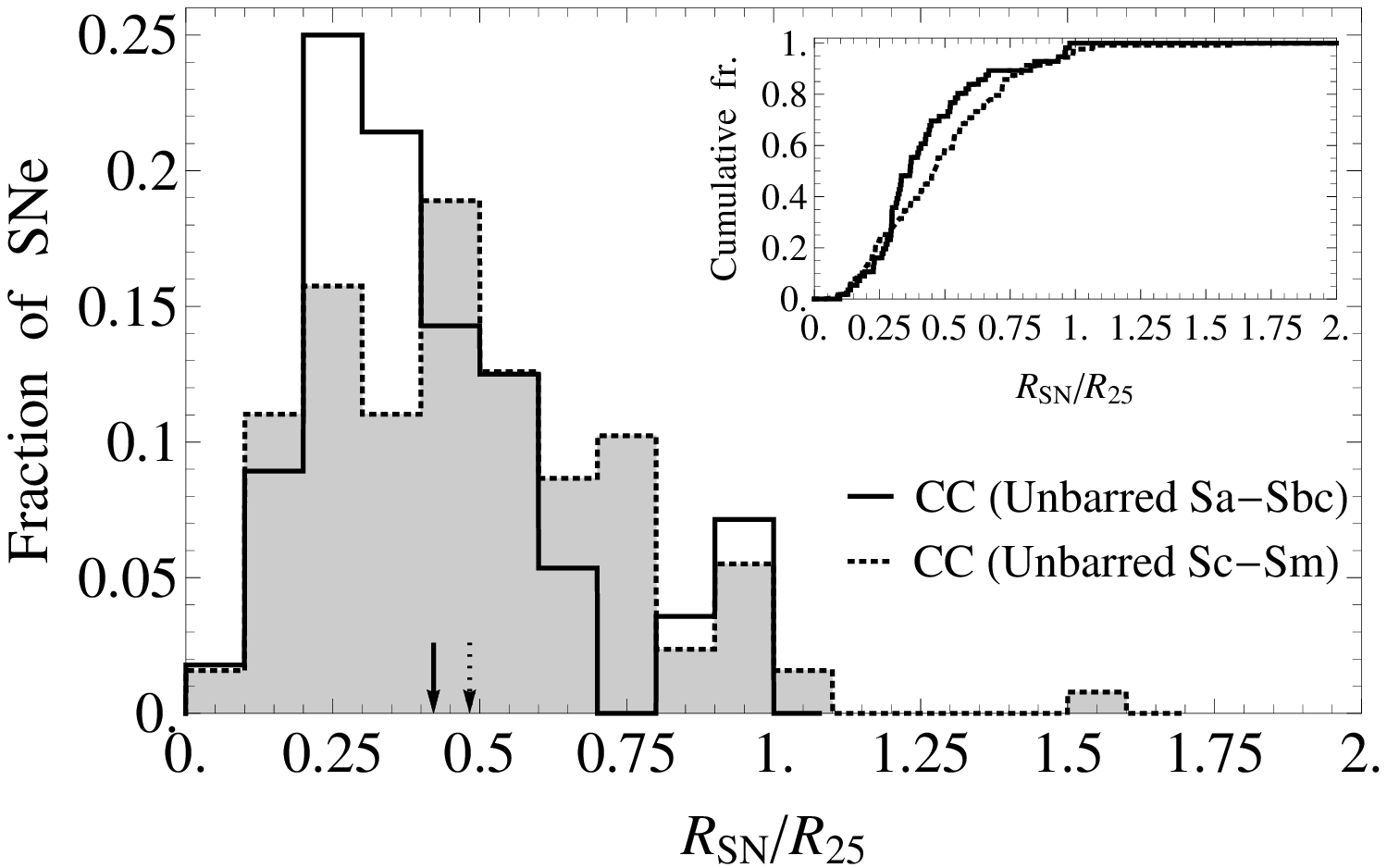}
\end{array}$
\end{center}
\caption{\footnotesize Left: the distributions of deprojected and normalized galactocentric radii
         of CC SNe in barred and unbarred Sa--Sbc hosts.
         Right: the distributions of CC SNe in unbarred Sa--Sbc and Sc--Sm galaxies.
         The insets present the cumulative distributions of SNe.
         The mean values of the distributions are shown by arrows.}
\label{hist_distr}
\end{figure}

\begin{enumerate}
\item In Sa--Sm galaxies, all CC and the vast majority of Type Ia SNe belong to the disc,
      rather than the bulge component.
      The result suggests that the rate of SNe Ia in spiral galaxies is dominated by
      a relatively young/intermediate progenitor population.
      This observational fact makes the deprojection of galactocentric radii of both
      types of SNe a key point in the statistical studies of their distributions.
\item The radial distribution of CC SNe in barred Sa--Sbc galaxies is not consistent
      with that of unbarred Sa--Sbc hosts (left-hand panel of Fig.~\ref{hist_distr}),
      while for Type Ia SNe the distributions are
      not significantly different.
      At the same time, the radial distributions of both
      Type Ia and CC SNe in Sc--Sm galaxies are not affected by bars.
      These results are explained by a substantial suppression of
      massive star formation in the radial range swept by strong bars of
      early-type barred galaxies \citep{2009A&A...501..207J,2015MNRAS.450.3503J}.
\item The radial distribution of CC SNe in unbarred Sa--Sbc galaxies is more centrally
      peaked and inconsistent with that in unbarred Sc--Sm hosts
      (right-hand panel of Fig.~\ref{hist_distr}). On the other hand, the radial distribution
      of Type Ia SNe in unbarred galaxies is not affected by host morphology.
      These results can be well explained by the distinct distributions of massive stars
      in the discs of early- and late-type spirals.
      In unbarred Sa--Sbc galaxies, star formation is more concentrated to the inner regions
      \citep[H$\alpha$ emission outside the nucleus, e.g. ][]{2009A&A...501..207J} and should thus be responsible
      for the observed radial distribution of CC SNe.
\item The radial distribution of CC SNe, in contrast to Type Ia SNe, is inconsistent with
      the exponential surface density profile,
      because of the central ($R_{\rm SN}/R_{25} \leq 0.2$) deficit of SNe.
      However, in the $R_{\rm SN}/R_{25}\in[0.2; \infty)$ range, the inconsistency
      vanishes for CC SNe in most of the subsamples of spiral galaxies.
      In the inner-truncated disc, only the radial distribution of CC SNe in barred early-type spirals
      is inconsistent with an exponential surface density profile, which
      appears to be caused by the impact of bars on the radial distribution of CC SNe.
\item In the inner regions of non-disturbed spiral hosts, we do not detect
      a relative deficiency of Type Ia SNe with respect to CC,
      contrary to what was found by other authors \citep[e.g.][]{1997ApJ...483L..29W}, who
      had explained this by possibly stronger dust extinction for Type Ia than for CC SNe.
      Instead, the radial distributions of both types of SNe are similar
      in all the subsamples of Sa--Sbc and Sc--Sm galaxies, which supports
      the idea that the relative increase of CC SNe in the inner regions of spirals found by the other
      authors is most probably due to the central excess of CC SNe in disturbed galaxies
      \citep[e.g.][]{2012MNRAS.424.2841H,2014MNRAS.444.2428H}.
\item As was found in earlier studies
      \citep[e.g.][]{2009A&A...503..137B,2009MNRAS.399..559A,2009A&A...508.1259H},
      the physical explanation for the more concentrated distribution of
      SNe Ibc with respect to SNe II in non-disturbed and unbarred spiral galaxies
      is that SNe Ibc arise from more metal-rich environments.
      The radial distributions of Types Ib and Ic SNe
      are sufficiently similar that the statistical tests fail to distinguish
      them with significance.
\end{enumerate}

\small  
%
\section*{Acknowledgments}   
%
{\footnotesize This work was supported by the RA MES State Committee of Science,
in the frames of the research project number 15T--1C129.
This work was made possible in part by a research grant from the
Armenian National Science and Education Fund (ANSEF) based in New York, USA.}

\bibliographystyle{aj}
\footnotesize
\bibliography{proceedings}

\end{document}